\documentclass{article}

\usepackage[preprint]{neurips_2024}

\usepackage{graphicx}
\usepackage{pgfplots}
\usepackage{mathtools}
\usepackage{hyperref}
\usepackage{dcolumn}
\usepackage{bm}
\usepackage{float}
\usepackage{amsmath}
\usepackage{algorithm}
\usepackage{algpseudocode}

\title{Parsimonious Hawkes Processes \\ for temporal networks modelling}

\author{Yuwei Zhu and Paolo Barucca \\ Department of Computer Science, University College London, WC1E 6BT London, United Kingdom}

\begin{document}

\maketitle

\begin{abstract}
Temporal networks are characterised by interdependent link events between nodes, forming ordered sequences of links that may represent specific information flows in the system. 
Nevertheless, representing temporal networks using discrete snapshots in time partially cancels the effect of time-ordered links on each other, while continuous time models, such as Poisson or Hawkes processes, can describe the full influence between all the potential pairs of links at all times.
In this paper, we introduce a continuous Hawkes temporal network model which accounts both for a community structure of the aggregate network and a strong heterogeneity in the activity of individual nodes, thus accounting for the presence of highly heterogeneous clusters with isolated high-activity influencer nodes, communities and low-activity nodes. 
Our model improves the prediction performance of previously available continuous time network models, and obtains a systematic increase in log-likelihood.
Characterising the direct interaction between influencer nodes and communities, we can provide a more detailed description of the system that can better outline the sequence of activations in the components of the systems represented by temporal networks.
\end{abstract}

\maketitle

\section{Introduction}

Networks are a fundamental representation of a system’s interactions. When a system presents temporal variability in the strength of the interaction between its elements then time-varying graphs,  \citep{masuda2016guide}, are a better representation of the system than static ones. Examples of dynamic networks are found in social networks \citep{newman2001structure}, physical proximity networks \citep{eagle2009inferring}, or financial and economic networks \citep{mazzarisi2020dynamic}, including input-output trade networks \citep{timmer2012world} or supply-chain networks \citep{pichler2023building}.  \\
Network theory has developed advanced tools for the understanding of the properties of static, i.e. time-independent, networks through the estimation of static ensembles of networks compatible with a given set of constraints on some relevant network quantities, e.g. node degree, node strength or both \citep{newman2018networks,cimini2019statistical}. 
More recently, a number of studies have been dealing with the explicit modelling of the temporal evolution of networks with discrete and continuous models, as distinguished by the way they represent the time evolution of links\citep{clemente2023temporal}. 
In particular, a broad range of network ensembles fall within the snapshots' representation \citep{masuda2016guide}, which aggregates the links between nodes into fixed time windows, providing a time-ordered collection of static networks.
The ensemble of temporal exponential random graph models (TERGMs), \citep{hanneke2006discrete}, is a prominent example of discrete snapshots' representation which leverages on the known results on static networks to analyse networks ordered in a temporal sequence of static networks. \\
Continuous models on the other hand use a continuous stochastic process for the generation of timestamped links between nodes, either modeled as instantaneous events happening at a given point time, thus identified with a unique timestamp, or as links with a finite life span, i.e. with a starting point and an ending point. 
In these kinds of models, the joint distribution of links over time is controlled by the underlying stochastic process and can reproduce properties of empirical sequences of timestamps, e.g. the distribution of the inter-event time between two subsequent links can reflect known phenomena of real systems such as burstiness \citep{lambiotte2013burstiness}. 
Going from a sequence of static networks to a more granular temporal network description can lead to disentangling causal effects in the system in a way that a snapshots' representation simply cannot capture due to the information loss which occurs with aggregation.  
Recent works on the use of continuous latent space dynamics for the evaluation of the likelihood of a given observed sequence of events have shown the potential of such approach for modeling realistic properties of financial \citep{bacry2015hawkes, bacry2018tick, jain2024limit, seabrook2022modelling, mercuri2024hawkes} and social systems \citep{arastuie2020chiphawkesprocessmodel}, and are being extended to also consider block-structured, non-linear, and long-memory network dependencies \citep{soliman2022multivariatecommunityhawkesmodel}. \\
In particular, the Community Hawkes Independent Pairs (CHIP) generative network model was proposed, \citep{arastuie2020chiphawkesprocessmodel}, to extend univariate Hawkes modeling to temporal networks, by grouping nodes into communities extracted from the static network of interactions obtained by aggregating links over the entire available time window.  
The Multivariate community Hawkes (MULCH) model, \citep{soliman2022multivariatecommunityhawkesmodel}, recently generalised the CHIP model by considering different modes of excitation, including reciprocity and several motifs, distinguishing between excitations coming from different blocks, hence making a more detailed use of the network clustering in the Hawkes process. 
Nevertheless, an important feature of networked systems is the emergence of heterogeneous profiles among nodes which lead to a broad distribution of the influence that each node can exercise on the network, as in the case of the rich-get-richer effect or in the presence of core-periphery structures\citep{holme2005core}. 
The degree heterogeneity of nodes, which emerges when links are aggregated over a sufficiently long time window, can reflect the variation of the activity level of the different elements of the system in the influence they exert on each other and over time, i.e. some elements of the system can be more active than others and they can be more active in some time periods and less in others. 
Further, high-activity nodes, here referred to as influencers, can greatly differ in the way they interact with the rest of the network over time and between each other. 
A block-representation of the interactions between the links of the temporal network can force different nodes in the same block to share the same interaction patterns with other nodes, hence potentially leading to an inaccurate description of the system when these nodes are particularly influential inside the network. Depending on the clustering method, nodes that belong to the same cluster may have a completely different interaction pattern in terms of directed influence towards other nodes and connections. 
\\
In this study, we introduce a continuous Hawkes temporal network model which accounts both for a community structure in the nodes of the temporal network and a strong heterogeneity in the influence of individual nodes. 
When we separate high-activity influencer nodes from community nodes, we improve the prediction accuracy with respect to previous models as well as outperforming them in terms of log-likelihood. 
In light of the inferred interactions between influencer nodes and communities, as estimated by our multivariate Hawkes model, we describe and interpret three benchmark temporal network datasets: Militarized Interstate Dispute (MID) data, reality mining data, and Enron mailing data. 
Then, we discuss the limitations of the present implementation of the algorithm, mainly in terms of the scalability of the model to higher dimensions for the multivariate Hawkes process and with respect to the set of temporal network features that might be used to better inform the influencers' selection and the clustering. Finally, we provide perspectives for future implementations of the model and for its potential applications. 

\section{Methods}
\subsection{Hawkes Processes}
Hawkes processes are self-exciting point processes, i.e. stochastic processes that model events occurring at random points in time. 
Self-excitation refers to the fact that if one event occurs at a given time, another event is more (or less) likely to occur soon after. 
This special property differentiates the Hawkes process from other related processes such as Poisson processes, and makes it effective for modelling many real-life scenarios, such as interactions on social media, or modelling volatility in finance \citep{jain2024limit,mastromatteo2011criticality}. 
The general equation for a $D$-variate Hawkes process \citep{hawkes1971spectra, bacry2016first} is as follows:
\begin{equation}
h_i(t) = \mu_i + \sum_{j=1}^{D} \int_{-\infty}^{t} \phi_{ij}(t - s)dN_j(s)
\end{equation}
where $N$ is the $D$-variate point process, $\lambda$ is the vector of the intensities characterising the probability of a new event for each component of the point process, $\mu$ is the vector of exogenous intensities, and  $\phi_{ij}(t)$ are commonly referred to as kernel functions, often taken to be an exponential or a power-law. 
Here, we will consider exponential kernels with hyperparameters $\alpha, \gamma$:
$$\phi_{ij}(x) = \alpha_{ij} e^{-\gamma_{ij} x}$$
In the following, we will mainly characterise and discuss systems based on the estimated values for $\alpha$ in their temporal network representation. 

\subsection{Multivariate community Hawkes (MULCH)}

We base our model on the Multivariate community Hawkes (MULCH), introduced by Soliman et al. \citep{soliman2022multivariatecommunityhawkesmodel}. The MULCH model adopts a community structure similar to the Stochastic Block Model (SBM) where instead of considering individual link-to-link interactions which can become intractable, i.e. of order $\mathcal{O}(n^4)$, we consider a simplified problem with $K$ communities and community-to-community interactions of order $\mathcal{O}(K^2)$. A particular node pair $(x, y)$, or link, can only influence other node pairs in their respective communities $(a, b)$ and every node pair in a specified block pair are assigned the same set of Hawkes parameters. Thus, we expect the behaviour of nodes assigned to the same community to be very similar. In the CHIP model, only one mode of excitation was considered - namely self-excitation. 
This means that for a given node pair $(x, y)$, an event occurring between these nodes at time $t$ will only influence the probability of an occurrence of an event between the same node pair $(x, y)$ from time t. Other related models such as (Blundell et al.) considered two modes of excitation, which included self-excitation and reciprocity i.e. $(x, y)$ can influence both $(x, y)$ and $(y, x)$, which generally makes sense in real life scenarios such as social media, where an interaction from user A to user B would often result in a response from user B back to user A. 
The MULCH model improved this further by considering six different modes of excitation, summarised in Table 1. Some of these modes capture more complex interaction motifs, for instance turn continuation, which captures the effect where $(x, y)$ will excite (or increase intensity) all connections between $x$ and all other nodes in the same block as $y$. To capture these new dynamics, the conditional intensity function between two nodes $(x, y)$ is captured by \eqref{eq:intensity}. The original self-excitation expression has now be replaced by a 6-term summation,

\begin{equation}
\lambda_{ij}(t) = \mu_{ab} + \sum_{\substack{(x, y) \in bp(a, b) \\ (x, y) \in  bp(b, a)}} \alpha_{ab}^{xy \rightarrow ij} \sum_{t_s \in T_{xy}} \gamma_{xy\rightarrow ij}(t-t_s)
\label{eq:intensity}
\end{equation}

accounting for each of the 6 modes of excitation. Notice that each node pair in a block pair $(a, b)$ share the same base intensity value (and also 6 excitation parameters). The gamma kernel function captures how the intensity of a given node pair change over time, commonly set to be the exponential kernel. This kernel decays the intensity function (when it is above the base intensity) back down to the baseline exponentially over time. The normalised form of the intensity function is used, see \eqref{eq:kernel}.

\begin{equation}
\gamma_{xy \rightarrow ij} (t) = \beta_{ab}^{xy\rightarrow ij} e^{-\beta_{ab}^{xy\rightarrow ij} t}
\label{eq:kernel}
\end{equation}

Whilst $\alpha$ and $\beta$ can both estimated via log-likelihood, it is common to assume a fixed $\beta$ value, or a hyper-parameter, and simply optimise for $\alpha$, simplifying the parameter estimation process. The choice of $\beta$ plays an important factor in parameter estimation; to account for this dependency, we use a weighted sum of different kernels. We can then associate each kernel with a multiplier $C_{i}$ that sum up to one, and tune these multipliers instead of optimizing for the $beta$ values directly.

\begin{equation}
\gamma_{xy \rightarrow ij} (t) = \sum_{q=1}^{Q} C_{ab}^{q} \beta_{q} e^{-\beta_{q} t}
\label{eq:weightedkernel}
\end{equation}

The new 'weighted' sum of kernels is shown in \eqref{eq:weightedkernel} - the different beta values, or timescales, are to be considered beforehand and are applied to all links in all communities. The parameter that differs between communities is restricted to the weighting parameters, $C_{ab}^{q}$ which is defined through parameter estimation. By increasing the number of $\beta$ kernels, we can refine the estimate with the cost of additional time complexity.

\begin{equation}
\lambda_{ij}(t) = \mu_{ab} + \sum_{\substack{(x, y) \in bp(a, b) \\ (x, y) \in  bp(b, a)}} \alpha_{ab}^{xy \rightarrow ij} \sum_{t_s \in T_{xy}} \sum_{q=1}^{Q} C_{ab}^{q} \beta_{q} e^{-\beta_{q} (t - t_s)}
\label{eq:fullintensity}
\end{equation}

The full log-likelihood equation, \eqref{eq:fullloglikelihood} reads

    \begin{equation}
    \begin{aligned}
    \ell_{a b}\left(\boldsymbol{\theta}_{a b} \mid \boldsymbol{Z}, \mathcal{H}_t\right)= & \sum_{Z_i=a, Z_j=b, i \neq j}\left\{-\mu_{a b} T-\sum_{\substack{(x, y) \in \operatorname{bp}(a, b) \\
    (x, y) \in \mathrm{bp}(b, a)}} \alpha_{a b}^{x y \rightarrow i j} \sum_{t_s \in T_{x y}} \sum_{q=1}^Q C_{a b}^q\left[\left(1-e^{-\beta_q\left(T-t_s\right)}\right)\right]\right. \\
    & \left.+\sum_{t_s \in T_{i j}} \ln \left[\mu_{a b}+\sum_{\substack{(x, y) \in \operatorname{bp}(a, b) \\
    (x, y) \in \mathrm{bp}(b, a)}} \alpha_{a b}^{x y \rightarrow i j} \sum_{q=1}^Q C_{a b}^q \beta_q R_{x y \rightarrow i j}^q\left(t_s\right)\right]\right\} \\
    & \text{where } \\
    & R_{x y \rightarrow i j}^q\left(t_s\right) = \sum_{\substack{t_r \in T_{xy} \\ t_r < t_s}} e^{-\beta_{q} (t_s - t_r)}
    \end{aligned}
    \label{eq:fullloglikelihood}
    \end{equation}

where $\boldsymbol{Z}$ represents the node membership and  $\mathcal{H}_t$ the full point process history. 
The MULCH model fitting process is as follows: node-to-node events are aggregated into a single event count over all time, which represents the adjacency matrix of the network. The initial node memberships are found using spectral clustering, then the model parameters are fitted for each block pair using the log-likelihood function and the non-linear L-BFGS-B optimizer. Soliman et al. also proposed a heuristic likelihood refinement procedure which allows the network model to ‘shuffle’ nodes around iteratively into other communities, which can improve upon the initial community assignment to reach a local optimum in terms of log-likelihood.

\begin{table}[h!]
    \centering
    \begin{tabular}{|l p{7cm}|} 
    \hline
    \textbf{Parameter} & \textbf{Excitation Type} \\
    \hline
    $\alpha_{ab}^{xy \rightarrow xy}$ & Self-excitation: continuation of event $(x, y)$. \\

    $\alpha_{ba}^{xy \rightarrow yx}$ & Reciprocal excitation: event $(y, x)$ taken in response to event $(x, y)$. \\

    $\alpha_{ab}^{xy \rightarrow xb}$ & Turn continuation: $(x, b)$ following $(x, y)$ to other nodes except for $y$ in the same block $b$. \\

    $\alpha_{ba}^{xy \rightarrow ya}$ & Generalized reciprocity: $(y, a)$ following $(x, y)$ to other nodes except $x$ in block $a$. \\
 
    $\alpha_{ab}^{xy \rightarrow ay}$ & Allied continuation: event $(a, y)$ following $(x, y)$ from other nodes except $x$ in block $a$. \\

    $\alpha_{ba}^{xy \rightarrow bx}$ & Allied reciprocity: event $(b, x)$ following $(x, y)$ from other nodes except $y$ in block $b$. \\
    \hline
    \end{tabular}
    \caption{6 different excitations considered in the MULCH model}
    \label{tab:excitations}
\end{table}

\subsection{Multivariate Influencers plus Communities Hawkes Model (MINCH)}
We present our MINCH model, which aims to enhance the clustering process of the MULCH model by distinguishing between communities and influential nodes. In MULCH, degree heterogeneity is partially dealt with by considering singular vectors for spectral clustering \citep{sussman2012universallyconsistentlatentposition} with row normalization \citep{Rohe2016CoclusteringDG} but we make the case that these additions are not enough to fully capture the contribution of outlier nodes in terms of level of activity and its specificity.  
Degree heterogeneity can strongly affect community detection by enforcing degree-assortative communities \citep{karrer2011stochastic,barucca2016disentangling}, and real systems often present a community structure on one side and isolated influential elements on the other. 
In such cases, degree heterogeneity can be accounted for by isolating and disentangling the contribution of high-degree nodes from the contribution to the system's dynamics coming from communities composed of more typical nodes. 
This could be achieved a number of different ways, in this paper we propose a proof-of-concept by implementing a simple heuristic that accounts for such outlier nodes in the initial node assignment phase, isolating nodes that exhibit different activity levels compared to the population.
This is important as very inactive or active nodes could potentially skew nodes in the same community to either have underestimated or overestimated parameters respectively. 
It is important to consider the physical interpretation behind this choice. For instance, in a scenario such as emails being sent between employees in a company, it is feasible to consider that a manager may send more emails to a wider range of recipients compared to regular employees, acting essentially as a super-spreader in the network, and this is an interpretation that we would like to capture in the network dynamics. 
Similar reasoning can be applied to nodes that do not interact much in the network. Consider an event matrix $E$, which contains tuples of events in the network $(i, j, t)$ from node $i$ to node $j$ at time $t$. $E = \{e_{0}, ..., e_{T}\}$. 
For a particular node $i$, we define the activity simply as the number of outgoing events from node i:

\begin{equation}
    a_x = \sum_{(i, j, t) \in E} \bm{1}_{\{x = i\}}
\label{eq:activities}
\end{equation}

We define the top $k$ nodes in terms of activity as hubs in our network, and low activity nodes are considered to be inactive if they are below a certain quantile threshold:

\begin{equation}
    \mathcal{A} = \{ i : a_i \geq a_{k} \}
\label{eq:hubs}
\end{equation}

\begin{equation}
    \mathcal{I} = \{ i : a_i \leq Q_p \}
\label{eq:inactive}
\end{equation}

Where $a_k$ is the k-th largest activity value, and $Q_p$ denotes the p-th quantile of activity values. These are defined as hyperparameters to our clustering algorithm. The hubs are defined each to be in their own clusters, due to the nature that each hub often acts independently and with its own different selection of clusters - multiple managers managing different teams - whereas the inactive nodes are simply assigned to a single cluster, which are expected to be fit with a weak baseline set of parameters.

\begin{algorithm}
\caption{Fitting MINCH}\label{alg:mulchfitting}
\begin{algorithmic}
\State \textbf{Input: } Number of clusters $k$, Event dictionary $E$, $n$-dimensional beta vector $\boldsymbol{\beta}$, end time of network $T$, low activity quantile threshold $Q_p$, Number of hubs in network $n_{hubs}$
\State \textbf{Output: } baseline intensity matrix $\boldsymbol{\mu}$, six $\alpha$ excitation matrices, $\{A_{s}, A_{r}, A_{c}, A_{gr}, A_{al} A_{alr}\}$, $n$ kernel weighting matrices $\{C_{1}, \dots, C_{n} \}$
\ForAll{Node pairs $i \neq j$}
    \State $N_{i,j} \gets$ number of events from $i$ to $j$ in $Y$
\EndFor
\State $\mathbf{v} \gets \textsc{SpectralCluster}(\mathbf{N}, k)$ \Comment{node memberships $\mathbf{v}$}
\State $\mathbf{a} \gets \text{activity values according to \eqref{eq:activities}}$
\State $t_{hubs} \gets n_{hubs} $ largests value in $\textbf{a}$
\ForAll{Nodes $i$}
    \If{$\mathbf{a}_{i} \geq t_{hubs}$}
        \State $\mathbf{v}_{i} \gets \text{new cluster assignment}$
    \ElsIf{$\mathbf{a}_{i} \leq Q_p$}
        \State $\mathbf{v}_{i} \gets \text{inactive cluster assignment}$
    \EndIf
\EndFor
\State Define log-likelihood function: 
       $\ell_{ab}(\boldsymbol{\theta}_{ab} \mid \boldsymbol{Z}, \mathcal{H}_t)$
\State Define the Jacobian of the log-likelihood with respect to the parameters $\boldsymbol{\theta}_{ab}$:
       \[
       \mathbf{J}_{\ell_{ab}}(\boldsymbol{\theta}_{ab} \mid \boldsymbol{Z}, \mathcal{H}_t) = 
       \begin{bmatrix}
       \frac{\partial \ell_{ab}}{\partial \theta_j}\ \text{for}\ \theta_j \in \boldsymbol{\theta}_{ab}
       \end{bmatrix}^{\top}
       \]
       where each element represents the partial derivative of $\ell_{ab}$ with respect to the parameter $\theta_j$.
\For{each Block a}
    \For{each Block b}
        \State $\mathcal{L} \gets \ell_{ab}(\boldsymbol{\theta}_{ab} \mid Z_i=a, Z_j=b, \mathcal{H}_t)$
        \State $\mathcal{J} \gets \mathbf{J}_{\ell_{ab}}(\boldsymbol{\theta}_{ab} \mid Z_i=a, Z_j=b, \mathcal{H}_t)$
        \State $n_a \gets |a|$ \Comment{Number of nodes in block a}
        \State $m \gets |a||b|$ \Comment{Node pairs in block pair $(a,b)$}
        \State $E_{ab} \gets \text{events in } E$ involving block pair $(a,b)$
        \State $\mathcal{E} \gets (E_{ab}, T, n_a, m, \boldsymbol{\beta})$
        \State $\mu_{ab}, \boldsymbol{\alpha}_{ab}, \mathbf{C}_{ab} \gets \textsc{LBGFSBOptimizer}(\mathcal{L}, \mathcal{J}, \mathcal{E})$
    \EndFor
\EndFor
\end{algorithmic}
\end{algorithm}

\section{Results}

In this section, we present the results of fitting MINCH to three different real-life network datasets. The raw data took the form of tabular data, where each row contained two node identifiers (sender, receiver) along with a timestamp when it occurred. Everything to do with the setup of the dataset was kept consistent with MULCH to ensure the reproducibility of the results and ensure a fair comparison. We use the mean test log-likelihood as the evaluation metric, holding out the last 20 percent of data (split by time) to use as the test set. Thus, we fit the model parameters on the first 80 percent of data and use these parameters to calculate the log-likelihood per event on the test set. Any previously unseen nodes are automatically assigned to the largest community.

\begin{table}[h!]
\centering
\begin{tabular}{ |p{4cm}|p{2cm}|p{2cm}|  }
 \hline
Dataset & MULCH & MINCH \\
 \hline
 Reality Mining   & -3.82 & \textbf{-3.79} \\
 Enron &   -5.13  & \textbf{-5.11}\\
 MID & -3.53 & \textbf{-3.51}\\
 \hline
\end{tabular}
\vspace{10pt}
\caption{Comparison of average test event log-likelihood between MULCH and MINCH models across different datasets. Lower values indicate better performance, with the best value for each dataset highlighted in bold.}
\label{tab:mulch_minch_comparison}
\end{table}

From the table of results \ref{tab:mulch_minch_comparison}, we see that MINCH provides a small, but systematic improvement in the test evaluation metric for the three datasets, demonstrating a higher performance in the model's ability to predict future events.

\clearpage
\section{MID case study}

\begin{figure}
    \centering
    \includegraphics[width=1\linewidth]{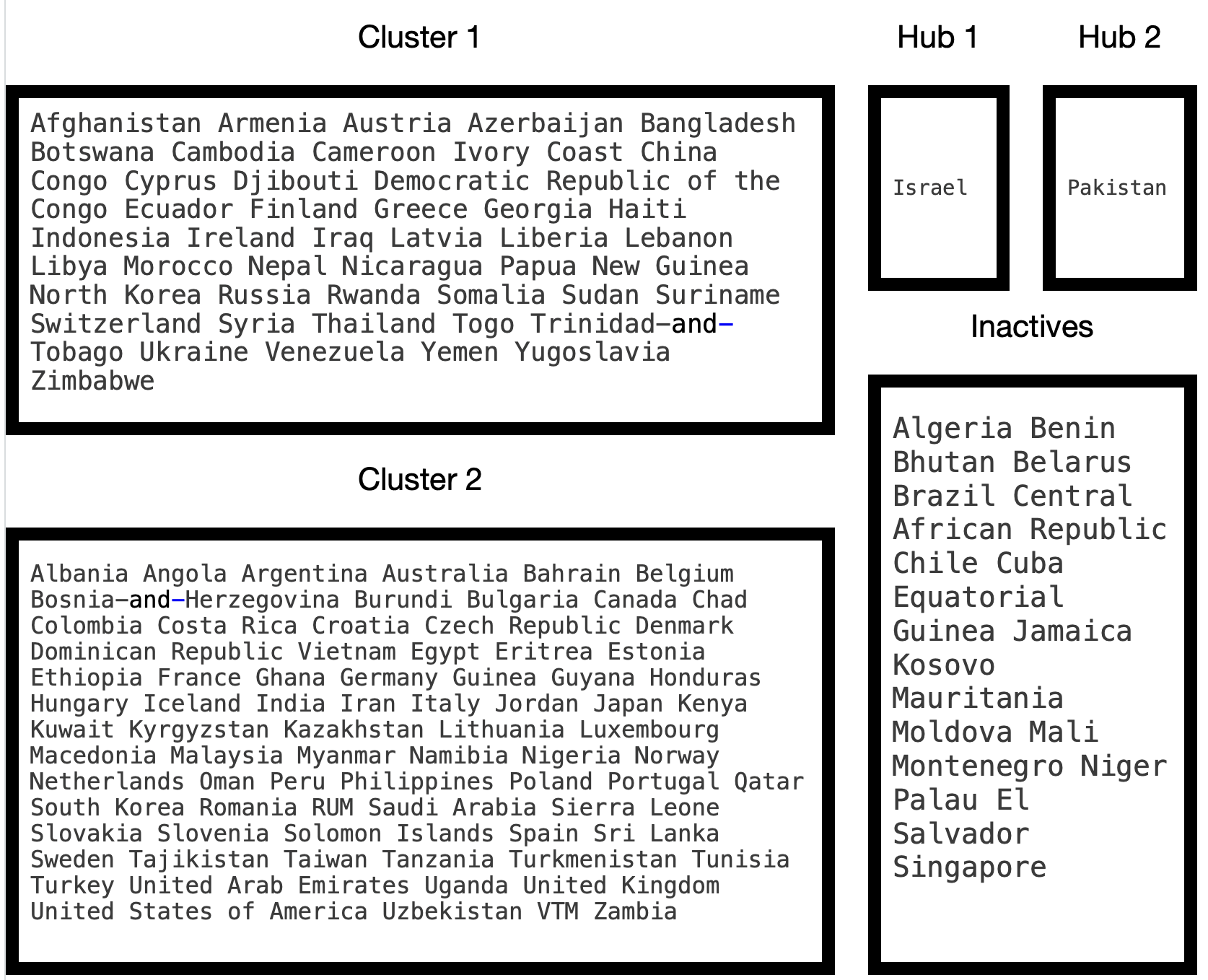}
    \caption{MID clusters produced by MINCH}
    \label{fig:MINCH}
\end{figure}

For the Militarized interstate disputes (MID) dataset we see that, excluding the outlier groups, the model has identified two main groups in the military disputes dataset. Cluster 2 has many similarities to block 1 identified by MULCH; this cluster contains 28 out of a possible 31 members of NATO, and includes numerous countries that are often considered allied with NATO member states, such as UAE, which participated in multiple military movements with NATO in the MID dataset from 1993-2014. 

The MINCH model has identified multiple 'inactive' states that have had one or less military movements during the period covered by MID, and identified Israel and Pakistan as the two hubs of this network. One block-to-block interaction that stands out is the strong link between cluster 1 and hub 1 (Israel), which had over 500 interactions during the 21 year period, which is comparable to the amount of interactions between clusters during the same time period. 
Interestingly, the MINCH model captures one of the most prevalent disputes during the period covered by the MID dataset. Cluster 1 contains countries commonly in dispute with Israel, including Cyprus, Iraq, Lebanon, Sudan, and Syria.

\begin{figure}
    \centering
    \includegraphics[width=1\linewidth]{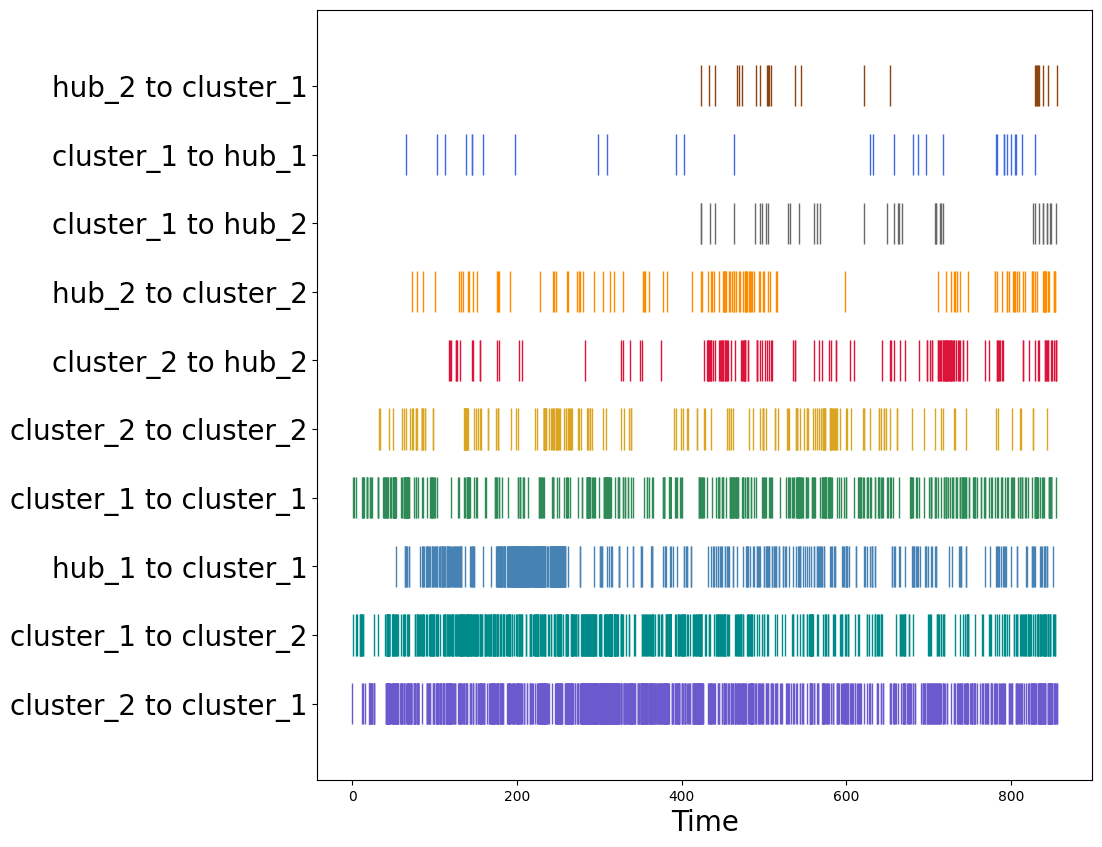}
    \caption{Event interactions between blocks over time}
    \label{fig:mid_raster}
\end{figure}

Another relevant set of interactions is between hub 2 (Pakistan) and cluster 2 (majority NATO or NATO allies), which is elevated, when taking into account the total number of node pairs. This time, the events between the two blocks are relatively equal. Again, we see this reflected in the baseline intensity and self-excitation parameters. This captures the array of conflicts between Iran-Pakistan and India-Pakistan that resulted in a quick succession of military disputes. Furthermore, there were multiple conflicts between NATO and Pakistan, especially between the USA. This is reflected in the reciprocal and turn continuation parameters from cluster 2 to hub 2.

\begin{figure}
    \centering
    \includegraphics[width=0.8\linewidth]{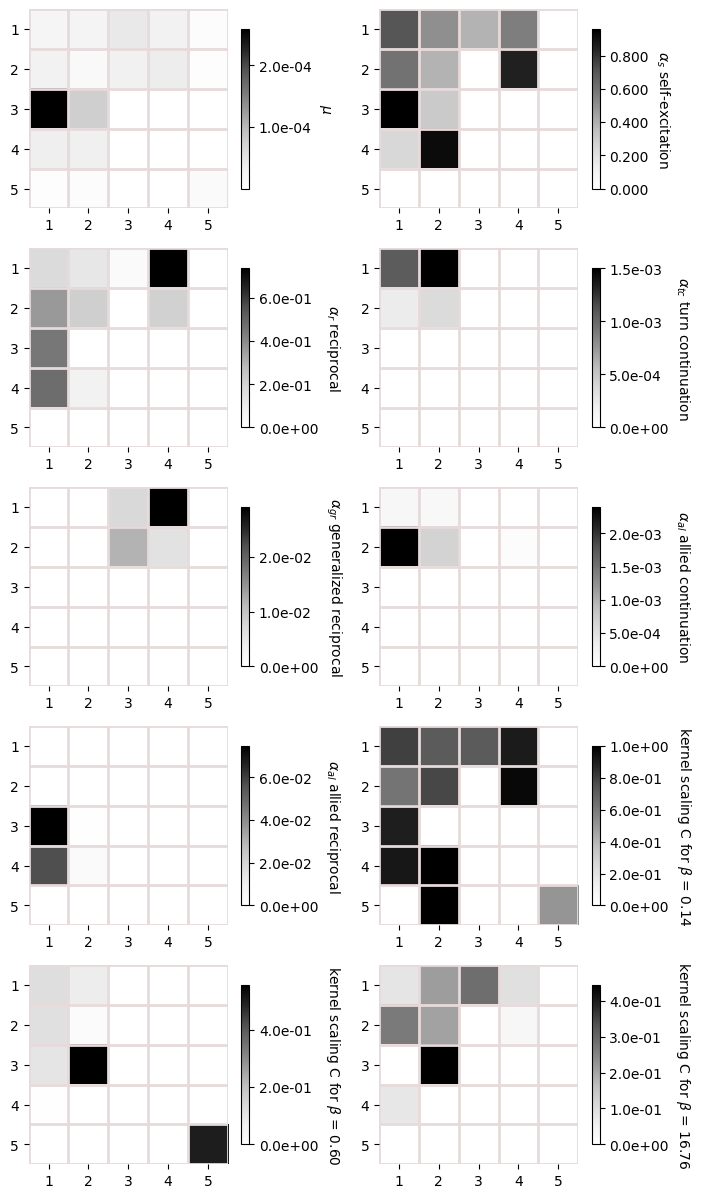}
    \caption{MID model parameters for 5 communities (1 : cluster 1, 2: cluster 2, 3 : hub 1, 4 : hub 2, 5 : inactive); from left to right, top to bottom: $\mu$ base intensity, $\alpha_{s}$ self-excitation, $\alpha_{r}$ reciprocal, $\alpha_{c}$ turn continuation, $\alpha_{gr}$ generalized reciprocity, $\alpha_{al}$ allied continuation, $\alpha_{ar}$ allied reciprocity, $\beta_{halfday}$ kernel scaling, $\beta_{2weeks}$ kernel scaling, $\beta_{2months}$ kernel scaling model parameters. Row i and column j in the grids above represent block i to block j parameters.}
    \label{fig:MIDparams}
\end{figure}

\begin{figure}
    \centering
    \includegraphics[width=1\linewidth]{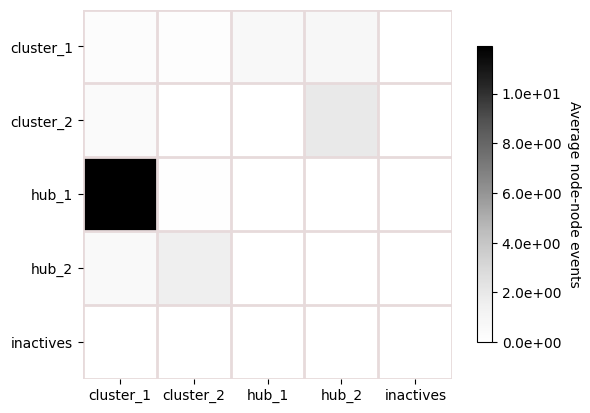}
    \caption{MID cluster - hub - inactive interactions. Aggregated events over time between different blocks in the network. Hubs refer to standalone nodes with large activity, and inactive nodes refer to nodes with a very low activity throughout the 21 year period. In particular, we see a large number of events directed from hub 1 to cluster 1, and in both directions between cluster 2 and hub 2.}
    \label{fig:MIDaggevents}
\end{figure}

\begin{figure}
    \centering
    \includegraphics[width=1\linewidth]{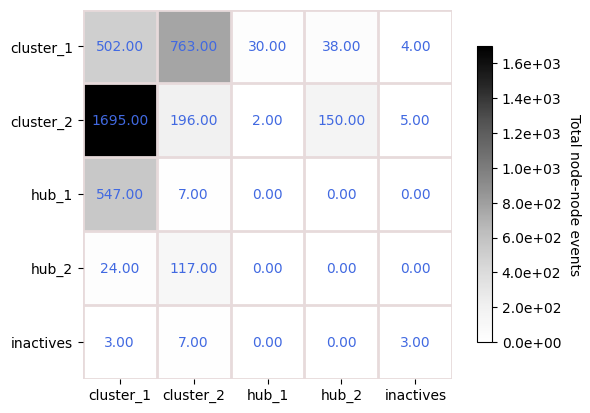}
    \caption{Total number of events recorded between blocks}
    \label{fig:enter-label}
\end{figure}

\clearpage

\section{Reality mining case study}

The Reality Mining dataset describes the in-person interactions between 100 students from Massachusetts Institute of Technology (MIT), which was collected over a period of 9 months. An event in this temporal network describes physical contact between two students.

\begin{figure}
    \centering
    \includegraphics[width=1\linewidth]{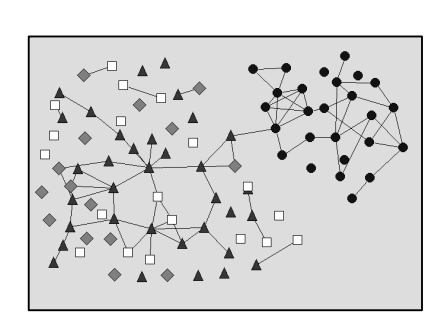}
    \caption{Underlying friendship networks based on collected data. \citep{Eagle2006RealityMS}
Circles represent incoming Sloan business school students. Triangles, diamonds and
squares represent senior students, incoming students, and faculty/staff/freshman at the Media Lab. }
    \label{fig:enter-label}
\end{figure}

The MINCH model identifies the correct number (2) of underlying clusters in the network as documented by the creators of dataset, who used the friendship data collected in the complimenting survey to graph the connections. The two clusters identified were: incoming Sloan business school students and the media lab faculty and students. However, the MULCH model has little preference between one or two clusters; showing that MINCH has been more adept at identifying the underlying structure from the data. Further, MINCH finds an interesting isolated interaction between two particular nodes in the network, hubs 2 and 3. Notice that the average events per node and total events between these two hubs are the same - indicating that these nodes have only ever communicated with each other for the duration of the experiment. This connection is also quite significant, accumulating a comparable amount of events to some of the main clusters. We can hypothesize that these two nodes could be two close friends, or a professor student mentorship. The friendship graph identified by the creators also shows multiple isolated two node connections, so this analysis is plausible.

\begin{figure}
    \centering
    \includegraphics[width=1\linewidth]{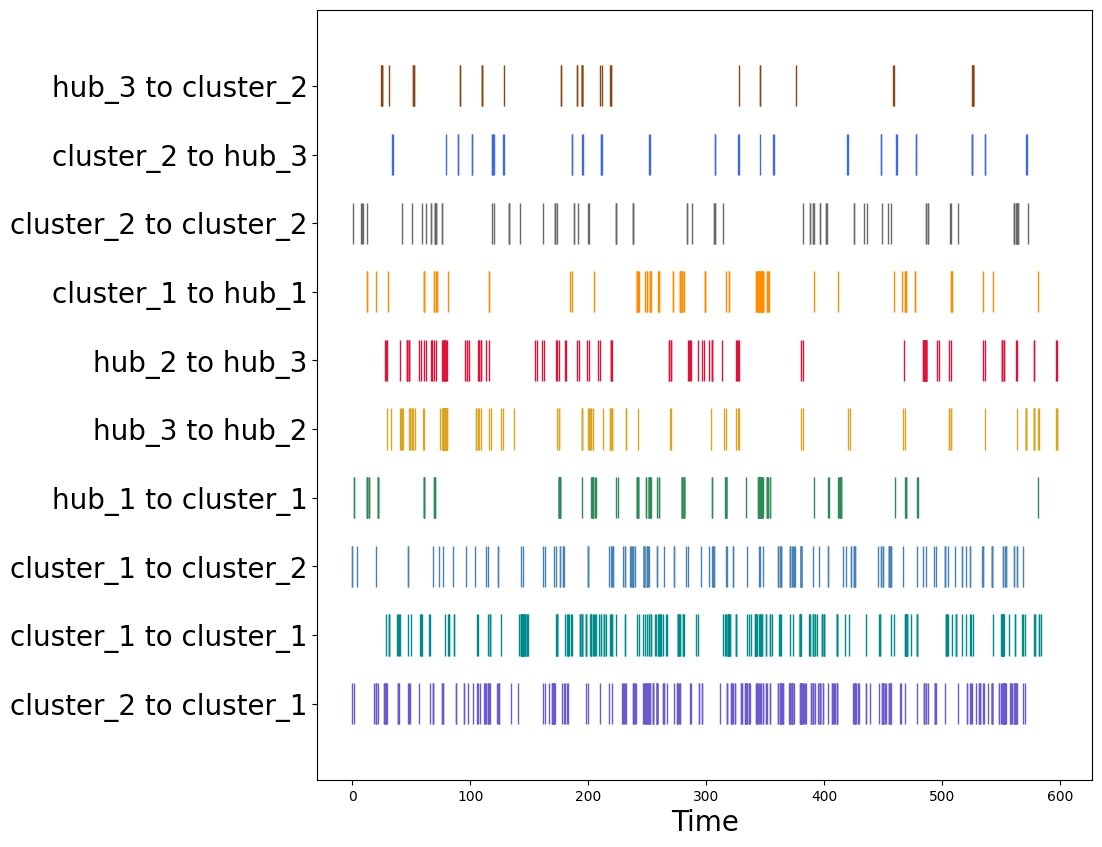}
    \caption{Reality mining block to block interactions}
    \label{fig:reality_raster}
\end{figure}

\begin{figure}
    \centering
    \includegraphics[width=1\linewidth]{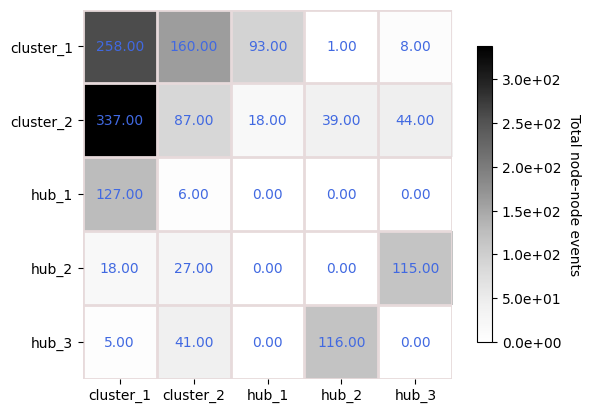}
    \caption{Block to block number of total interactions}
    \label{fig:enter-label}
\end{figure}

\begin{figure}
    \centering
    \includegraphics[width=1\linewidth]{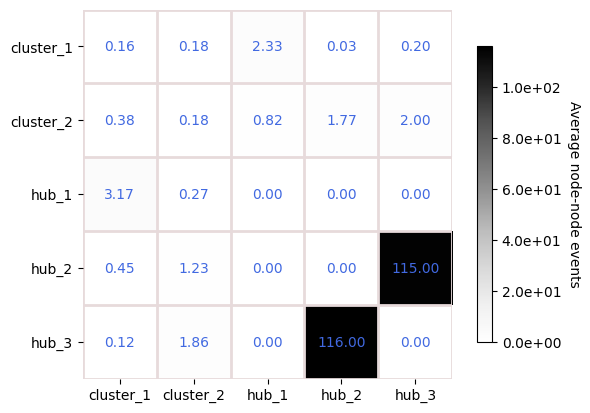}
    \caption{Average events per connection}
    \label{fig:avg-event-reality}
\end{figure}

\begin{figure}
    \centering
    \includegraphics[width=.8\linewidth]{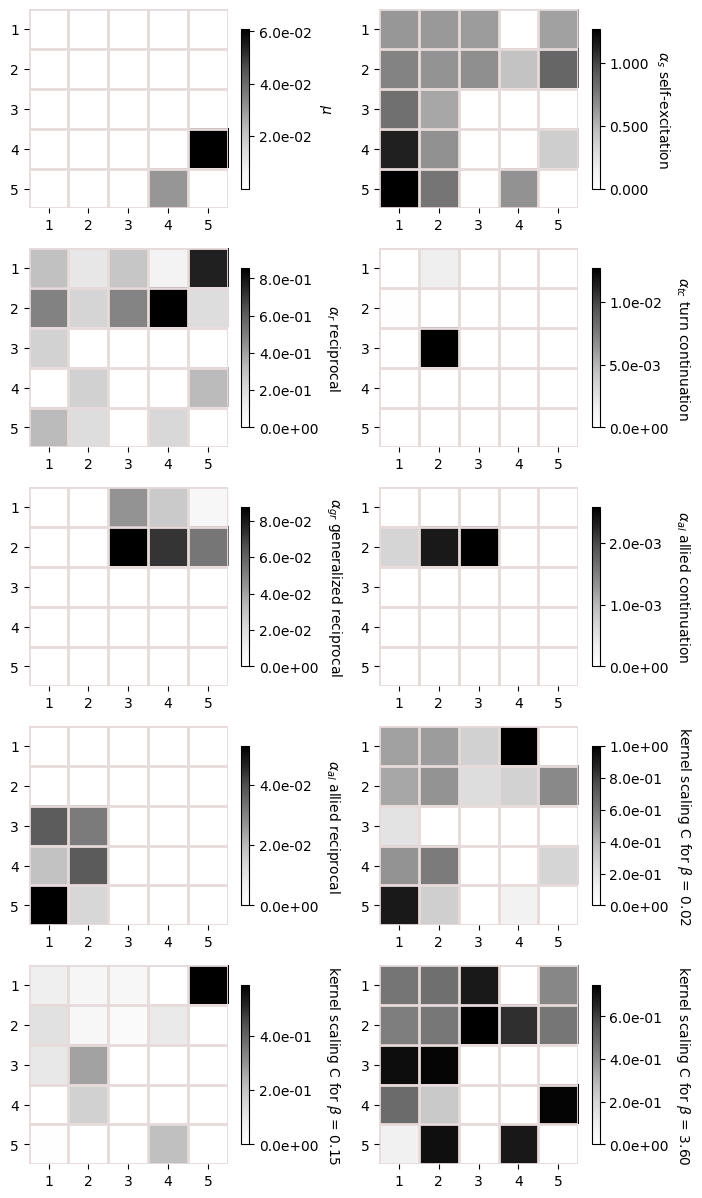}
    \caption{Reality mining model parameters for 5 communities (1 : cluster 1, 2: cluster 2, 3 : hub 1, 4 : hub 2, 5 : hub 3); from left to right, top to bottom: $\mu$ base intensity, $\alpha_{s}$ self-excitation, $\alpha_{r}$ reciprocal, $\alpha_{c}$ turn continuation, $\alpha_{gr}$ generalized reciprocity, $\alpha_{al}$ allied continuation, $\alpha_{ar}$ allied reciprocity, $\beta_{halfday}$ kernel scaling, $\beta_{2weeks}$ kernel scaling, $\beta_{2months}$ kernel scaling model parameters. Row i and column j in the grids above represent block i to block j parameters.}
    \label{fig:enter-label}
\end{figure}

\section{Enron dataset case study}
This dataset was collected and prepared by the CALO Project (A Cognitive Assistant that Learns and Organizes). It contains data from about 150 users, mostly senior management of Enron, organized into folders. The corpus contains a total of about 0.5M messages. In this case, MINCH simply identifies one main cluster, and two hubs that both communicate regularly with the main cluster yet displaying slight differences in interaction patterns. 

\begin{figure}
    \centering
    \includegraphics[width=0.8\linewidth]{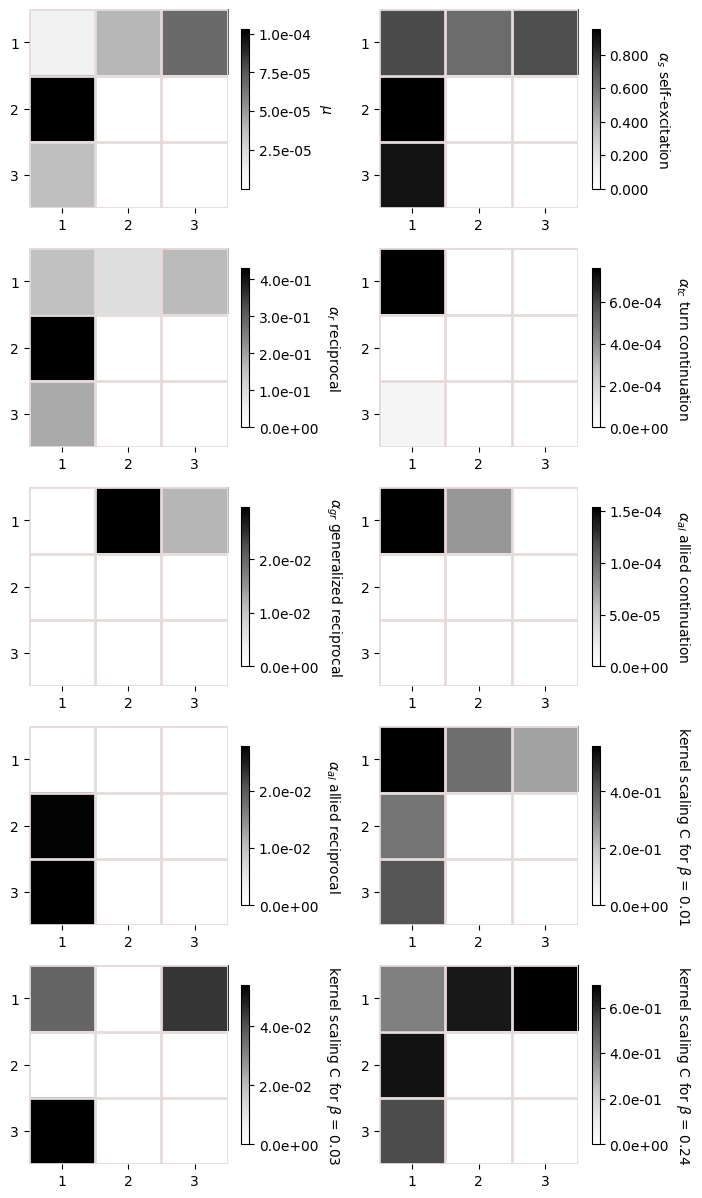}
    \caption{Enron model parameters for 3 communities (1 : cluster 1, 2: hub 1, 3 : hub 2); from left to right, top to bottom: $\mu$ base intensity, $\alpha_{s}$ self-excitation, $\alpha_{r}$ reciprocal, $\alpha_{c}$ turn continuation, $\alpha_{gr}$ generalized reciprocity, $\alpha_{al}$ allied continuation, $\alpha_{ar}$ allied reciprocity, $\beta_{halfday}$ kernel scaling, $\beta_{2weeks}$ kernel scaling, $\beta_{2months}$ kernel scaling model parameters. Row i and column j in the grids above represent block i to block j parameters.}
    \label{fig:enter-label}
\end{figure}

\begin{figure}
    \centering
    \includegraphics[width=1\linewidth]{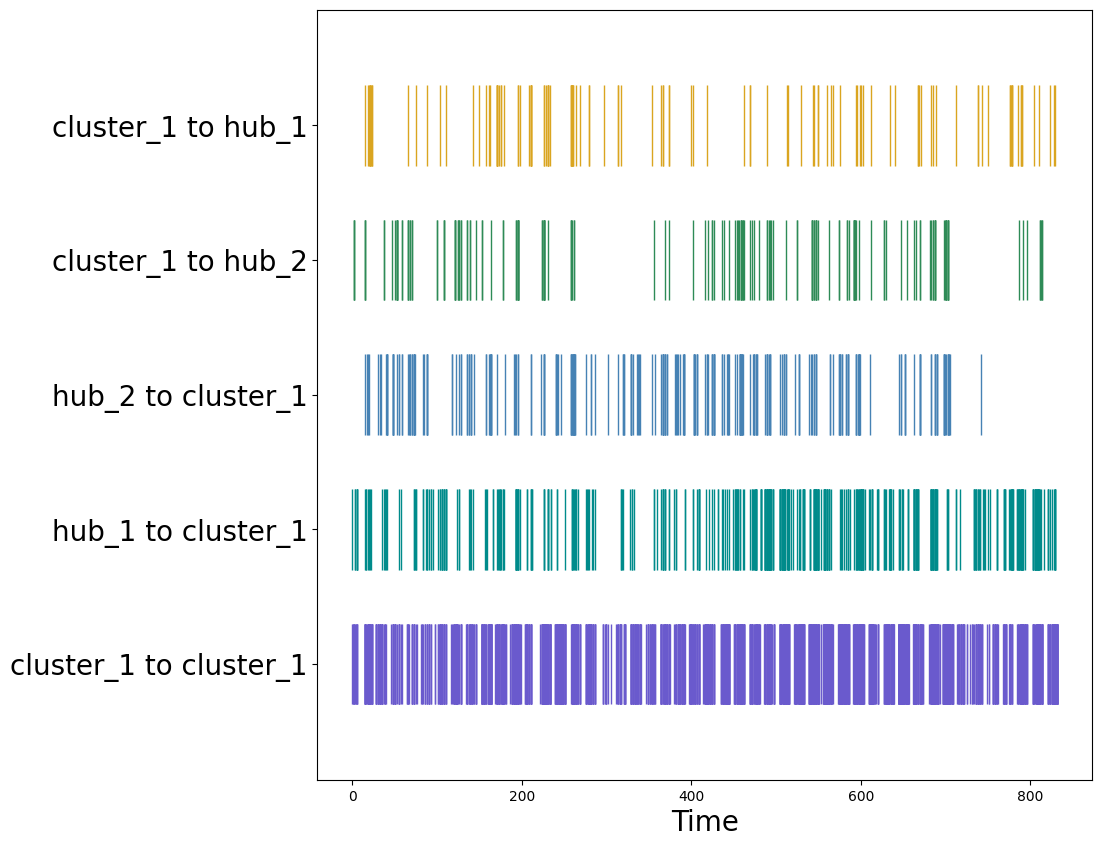}
    \caption{Enron block to block interactions}
    \label{fig:enter-label}
\end{figure}

\begin{figure}
    \centering
    \includegraphics[width=1\linewidth]{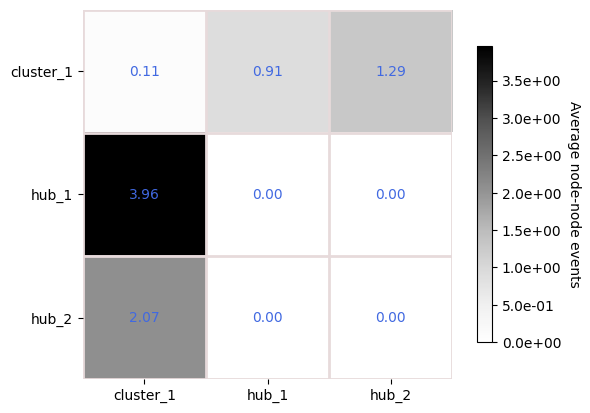}
    \caption{Average events per connection}
    \label{fig:enter-label}
\end{figure}

\begin{figure}
    \centering
    \includegraphics[width=1\linewidth]{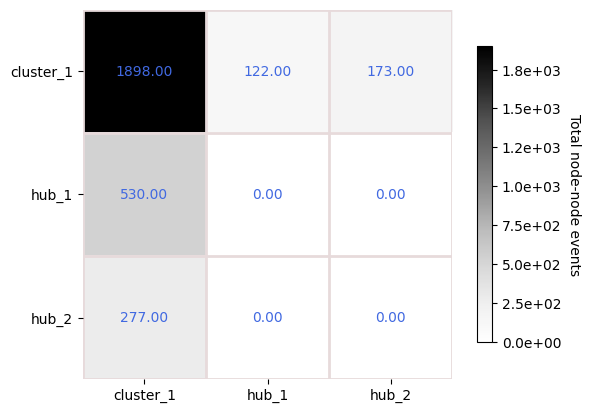}
    \caption{Total outgoing events per block}
    \label{fig:enter-label}
\end{figure}

\section{Discussion}
In this study we have shown how parsimonious Hawkes network models, such as MINCH, have a sufficient level of scalability and manageable time-complexity to be estimated on real temporal networks and can combine a rich time-evolution representation with a good explanatory power that clarifies the interaction between influential elements of the systems and communities of nodes. 
We have used both activity features and network clustering to provide a limited number of clusters and hubs, improving the explanatory power whilst not sacrificing on time complexity. 
We have shown the explanatory power of this approach explicitly in three use cases, MID dataset, reality mining, and in the Enron mailing dataset. 
This study shows promising results considering the simple and limited set of temporal network features used for distinguishing between influencers and clusters in the temporal network and also considering the simple fine-tuning for the kernel functions, which could also be carried out with deep machine learning architectures. 
Further, point process modeling could be applied to finite-time links representations, where links are no longer modeled by single events but as pairs of starting and ending events. Moreover, this activity-informed clustering with parsimonious multivariate Hawkes process modeling could be extended to temporal hyper-graphs, where events are represented by timestamped hyper-links. 

\clearpage

\bibliographystyle{plainnat}
\bibliography{temporalnets}

\end{document}